\documentclass[aps,pra,amsmath,amssymb,twocolumn,superscriptaddress,floatfix]{revtex4-2}%

\usepackage{xspace} % correct spacing 
\usepackage{graphicx}
\usepackage{bm}% bold math
\usepackage{xcolor}
\usepackage{color}

\begin{document}

\title{Ion-selective scattering studied by the variable-energy electron irradiation of Ba$_{0.2}$K$_{0.8}$Fe$_2$As$_2$ superconductor}

\author{Kyuil Cho}
\affiliation{Ames National Laboratory, Ames, IA 50011, USA}
\affiliation{Department of Physics, Hope College, Holland, MI 49423, USA}
\email{cho@hope.edu}
% Kyuil Cho: https://orcid.org/0000-0003-2111-6355

\author{M.~Ko\'nczykowski}
\affiliation{Laboratoire des Solides Irradi{\'e}s, CEA/DRF/IRAMIS, {\'E}cole Polytechnique, CNRS, Institut Polytechnique de Paris, F-91128 Palaiseau, France}
% Marcin Konczykowski: https://orcid.org/0000-0003-3376-5635

\author{M. A. Tanatar}
\affiliation{Ames National Laboratory, Ames, IA 50011, USA}
\affiliation{Department of Physics \& Astronomy, Iowa State University, Ames, IA 50011,	USA}
% Makariy Tanatar: https://orcid.org/0000-0003-2129-9833

\author{I. I. Mazin}
\affiliation{Department of Physics \& Astronomy, George Mason University, Fairfax, VA 22030, USA}
% Igor Mazin: https://orcid.org/0000-0001-9456-7099

\author{Yong Liu}
\affiliation{Ames National Laboratory, Ames, IA 50011, USA}
\affiliation{Crystal Growth Facility, Institute of Physics, {\'E}cole Polytechnique F{\'e}d{\'e}rale de Lausanne, CH-1015 Lausanne, Switzerland}
% Yong Liu: https://orcid.org/0000-0001-5161-2005

\author{T. A. Lograsso}
\affiliation{Ames National Laboratory, Ames, IA 50011, USA}
% Thomas Lograsso: https://orcid.org/0000-0002-8441-5320

\author{R. Prozorov}
\affiliation{Ames National Laboratory, Ames, IA 50011, USA}
\affiliation{Department of Physics \& Astronomy, Iowa State University, Ames, IA 50011,	USA}
% Ruslan Prozorov: https://orcid.org/0000-0002-8088-6096

\date{\today}

\begin{abstract}
Low-temperature variable-energy electron irradiation was used to induce non-magnetic disorder in a single crystal of hole-doped iron-based superconductor, Ba$_{1-x}$K$_x$Fe$_2$As$_2$, $x=$0.80. To avoid systematic errors, the beam energy was adjusted non-consequently for five values between 1.0 and 2.5 MeV, whence sample resistance was measured in-situ at 22 K. For all energies, the resistivity raises linearly with the irradiation fluence suggesting the creation of uncorrelated dilute point-like disorder (confirmed by simulations). The rate of the resistivity increase peaks at energies below 1.5 MeV. Comparison with calculated partial cross-sections points to the predominant creation of defects in the iron sublattice. Simultaneously, superconducting $T_c$, measured separately between the irradiation runs, is monotonically suppressed as expected since it depends on the total scattering rate, hence total cross-section, which is a monotonically increasing function of energy. Our work confirms experimentally an often-made assumption of the dominant role of the iron sub-lattice in iron-based superconductors.
\end{abstract}
\maketitle

\section{Introduction}

Response of superconductivity to the impurities and defects provides a useful tool to study the pairing mechanism of superconductors~\cite{ Balatsky2006RMP, Alloul2009RMP_defect_cuprates, LiWu2016SST_review_FeSC_chem_impurity}. The isotropic $s-$wave paring state of conventional Bardeen-Cooper-Schrieffer (BCS) superconductors is robust against non-magnetic scattering. This statement is known as Anderson theorem~\cite{Anderson1959JPCS}. However, in the case of paramagnetic impurities, scattering involves simultaneous flipping of the spins of impurity and conduction electron, destroying singlet Cooper pairs. Thus, according to the Abrikosov and Gor’kov theory ~\cite{AbrikosovGorkov1961JETP}, conventional BCS superconductivity is suppressed and is destroyed at the finite critical value of the magnetic dimensionless scattering rate, $\Gamma =\hbar /\left( 2\pi k_BT_{c0}\tau \right)\approx0.14$. In the cases of the anisotropic or multiband superconducting order parameters, even nonmagnetic scattering is pair-breaking and leads to a suppression of $T_c$ \cite{Openov1998PRB_impurity,KoganProzorovMishra2013PRB}. 

Traditionally, chemical doping and alloying are used to induce extra scattering~\cite{Finnemore1965PR}. However, in addition to changing scattering rate, these cause changes in the electronic band structure, Fermi energy level, and build internal ``chemical pressure", all of which affect measurable properties.

Particle irradiation is an alternative way to generate scattering centers, and it has been intensively used to investigate the properties of materials. Depending on the choice of particles, the character of induced scattering centers varies from point-like defects, mostly vacancies (electron irradiation) ~\cite{Giapintzakis1994PRB_YBCO_e-irr, Cho2016ScienceAdvances_BaK122_e-irr, Cho2018SST_review_e-irr, Cho2018NatComm_NbSe2, Cho2022PRB_YBCO_e-irr}, to dendritic clusters (proton irradiation)~\cite{Nakajima2010PRB_BaCo122_proton_irr, Taen2013PRB_BaK122_proton_irr, Smylie2016PRB_BaP122_proton_irr, Moroni2017PRB_proton-irr, Sylva2018SST_FeSe_thinfilm_proton-irr, Torsello2022IEEE_FeSe_thinfilm_proton-irr}, and to columnar defects (heavy-ion irradiation)~\cite{KonczykowskiPRB1991_HeavyIon_YBCO, Nakajima2009PRB_HeavyIon, Prozorov2010PRB_HeavyIon, MurphyProzorov2013PRB_heavy_ion}. Furthermore, if the energy of the projectile particles varies, the character of defects generated also changes accordingly since the scattering dynamics significantly vary with the energy  \cite{Damask1963PointDefectsInMetals,Thompson1969}. 

In this contribution, we use variable-energy electron irradiation to experimentally determine which ions contribute most into the scattering rate in iron-based superconductors, thus testing the models of electronic conductance in these materials. We chose Ba$_{1-x}$K$_{x}$Fe$_2$As$_2$ as one of the most intensively studied among the iron-based superconductors ~\cite{Dai2015RMP_revew_FeSCs_Neutron, Fernandes2017RPP_review, Cho2018SST_review_e-irr}. Here, superconductivity exists starting from $x=$ 0.16 and extends all the way to $x=$ 1. The abrupt change in the superconducting gap structure around $x=$0.7 was attributed to the Lifshitz transition ~\cite{Xu2013PRB_BaK122, RichardDing2015JPCM_APRES_IBS, Cho2016ScienceAdvances_BaK122_e-irr}. At low $x$, superconductivity coexists with long range magnetic order ~\cite{Avci2012PRB}. To avoid the influence of the magnetic phase and enable in-situ resistivity measurements, performed at a fixed 22 K in our setup, we chose the overdoped compound with $x=$ 0.8 with a convenient $T_{c,onset} = 20.2$ K.

\section{Materials and Methods}

\begin{figure}
\includegraphics[width=8.5 cm]{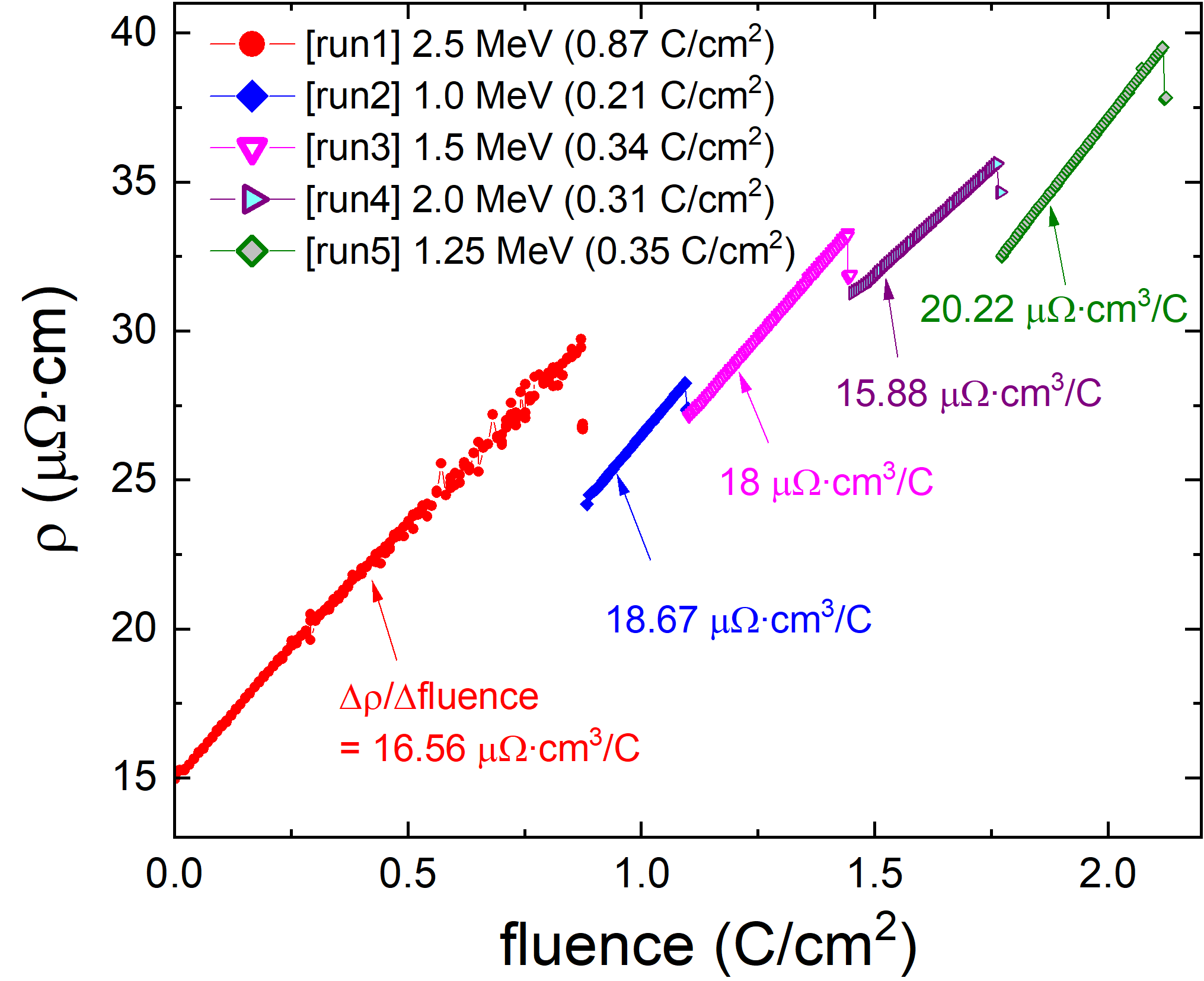}%
\caption{Fluence dependence of resistivity of Ba$_{0.2}$K$_{0.8}$Fe$_2$As$_2$ single crystal measured in-situ in irradiation chamber during electron irradiation. The sample was sitting in liquid hydrogen environment at a temperature around $T$ = 22 K. Five irradiation runs were conducted in order (sections of broken line in the figure). After each irradiation, the sample was taken out of the irradiation chamber for characterization (as shown in Fig.~\ref{fig3} (a)) and returned for the next irradiation. Sample thermal cycle to room temperature resulted in partial disorder annealing and slight resistivity decrease compared to the value at the end of the previous run.}
\label{fig1}
\end{figure}

\begin{figure}
\includegraphics[width=8.5 cm]{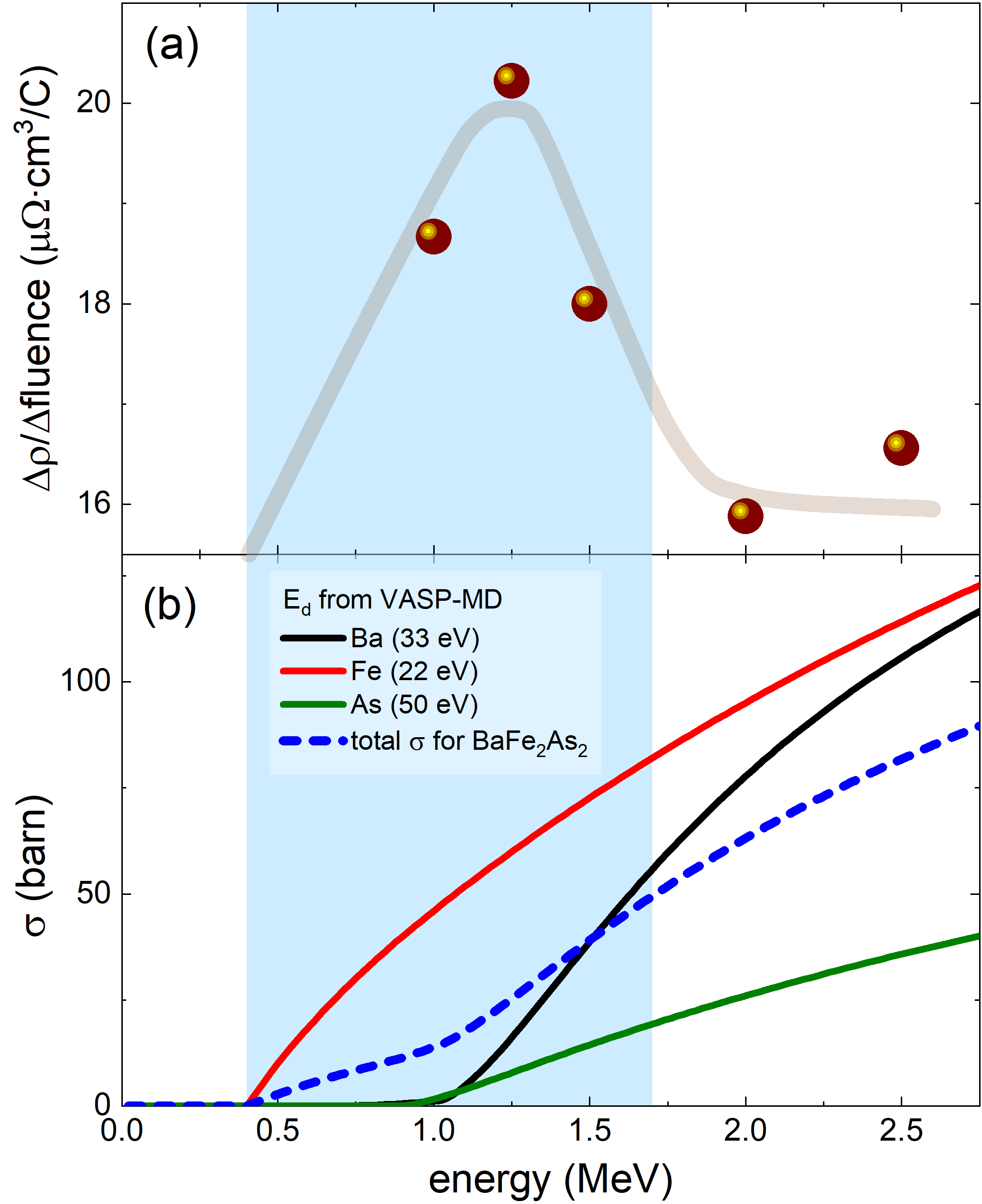}%
\caption{(a) Rate of the in situ resistivity increase with the fluence (slope of the lines segments in Fig.~\ref{fig1}) plotted as a function of the energy of electron beam. $\Delta \rho$ is the increase of resistivity during each irradiation run and $\Delta fluence$ is the total fluence for that irradiation. The lower energy irradiations show larger rate of in-situ resistivity per fluence. (b) Energy dependent scattering cross-sections for Ba, Fe, and As are calculated using the displacement energy (E$_{d}$) = 33 eV (Ba), 22 eV (Fe), and 50 eV (As), which were calculated from VASP-MD simulation. Total cross-section for BaFe$_2$As$_2$ is plotted as a dashed line.}
\label{fig2}
\end{figure} 

Single crystals of Ba$_{0.2}$K$_{0.8}$Fe$_2$As$_2$ were grown by using an inverted temperate gradient method with the starting materials - Ba and K lumps, and Fe and As powders. Details of the growth method can be found elsewhere \cite{Liu2013PRB, LiuProzorovLograsso2014PRB, Cho2016ScienceAdvances_BaK122_e-irr}. Resistivity measurements were performed in a standard four-probe configuration. Typical dimension of the samples are (1-2) $\times$ 0.5 $\times$ (0.02-0.1) mm$^3$. Silver wires of 50 $\mu$m diameter were
soldered to the sample to provide electrical contacts \cite{Tanatar2010SST}. The sample was mounted on a Kyocera chip over a hole of about 5 mm diameter at the center. The Kyocera chip was transferred to the irradiation chamber filled with liquid Hydrogen providing efficient cooling down to 22~K. A Faraday cup placed behind the chamber enabled accurate measurement of the fluence during irradiation. The electron irradiation was performed at the SIRIUS Pelletron facility of the Laboratoire des Solides Irradies at the Ecole Polytechnique in Palaiseau, France. The energy of electron beam was varied from 1.0 MeV to 2.5 MeV. The acquired irradiation dose is conveniently measured in C/cm$^2$, where 1 C/cm$^2$ = 6.24 × 10$^{18}$ electrons/cm$^2$. After irradiation, the sample in Kyocera chip is transferred to another set-up for temperature dependent resistivity measurement.

\section{Results and Discussion}

\begin{figure}
\includegraphics[width=8.5 cm]{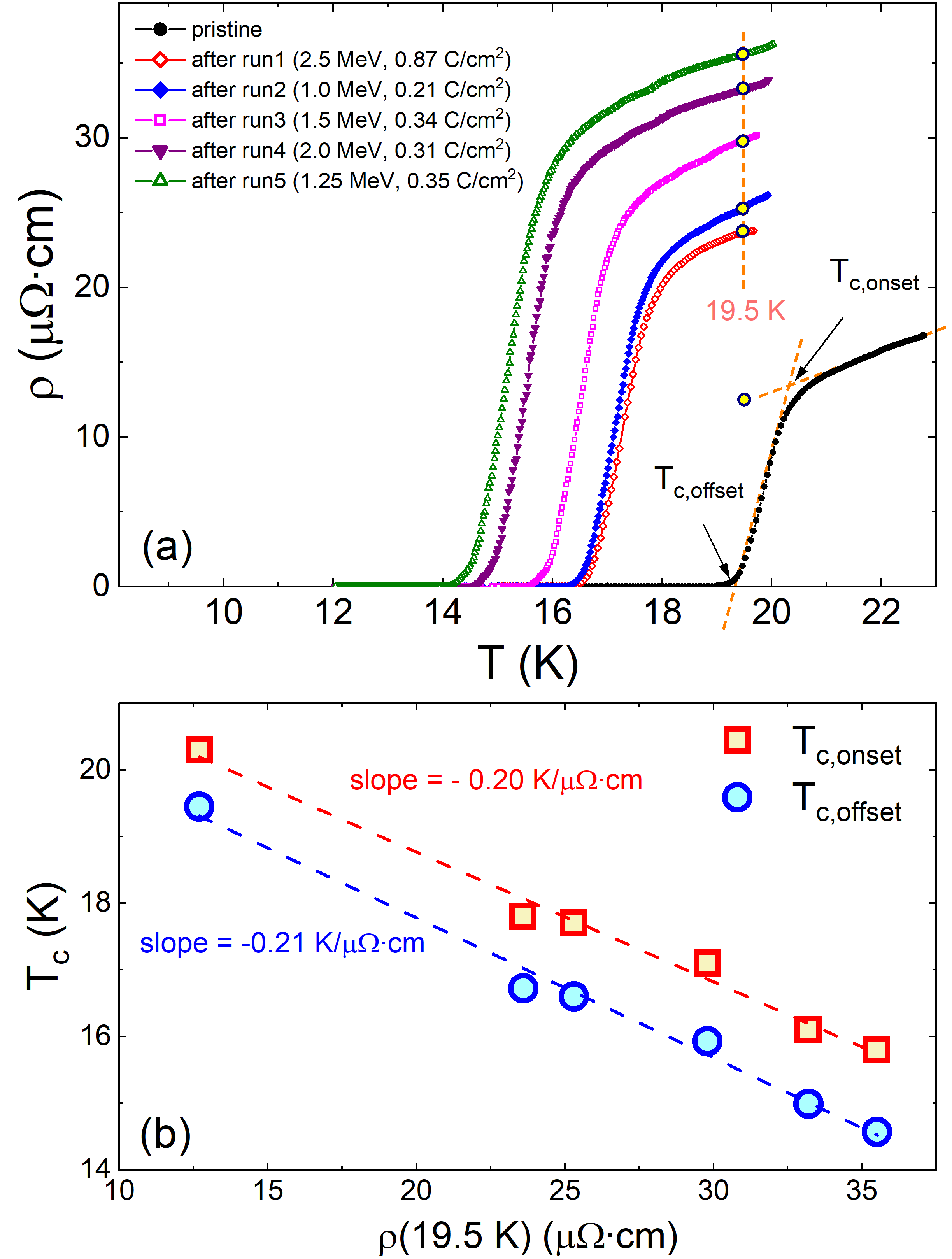}%
\caption{(a) Temperature-dependent resistivity measured after each run of irradiations. The sample was removed from irradiation chamber and transferred to a separate set-up to measure temperature-dependent resistivity. The normal state resistivity at 19.5 K (blue dotted line) is used as a parameter to indicate the amount of impurities generated upon irradiation. The definitions of the $T_{c,onset}$ and $T_{c,offset}$ are shown as red dotted lines. (b) $T_c$ versus resistivity at $T$ = 19.5 K. $T_{c}$ decreases at the rate of - 0.20 K/$\mu \Omega cm$ ($T_{c,onset}$) and -0.21 K/$\mu \Omega cm$ ($T_{c,offset}$).}
\label{fig3}
\end{figure} 

\begin{figure}
\includegraphics[width=8.5 cm]{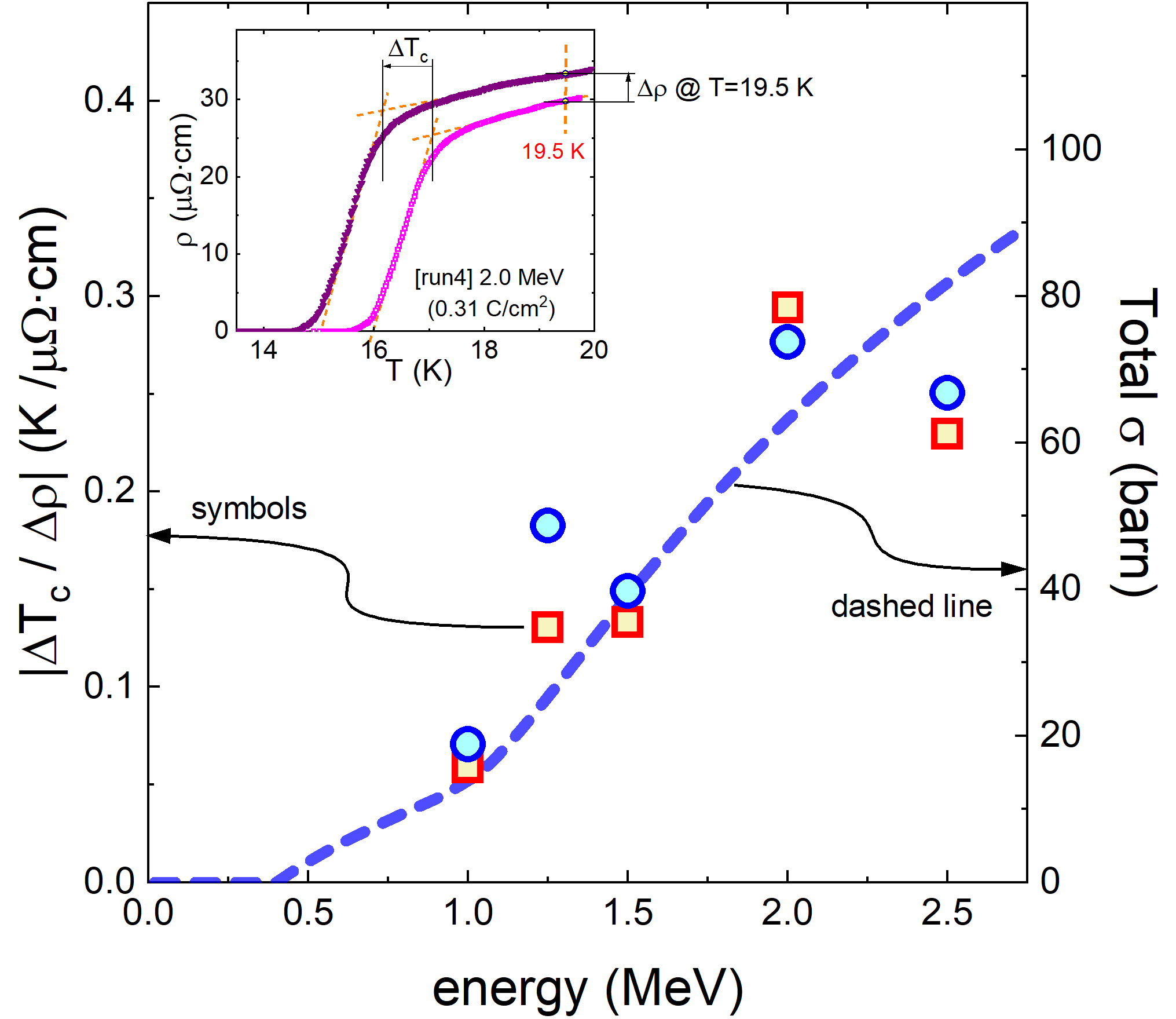}%
\caption{Normalized $T_c$ suppression rate ($|\Delta T_c / \Delta \rho|$) as a function of energy, calculated from the data in Fig.~\ref{fig3} (a). The inset shows the definition of $\Delta T_c$ and $\Delta \rho$ for a particular `run4' (2 MeV irradiation, 0.31 C/cm$^2$) as an example. The fact that the suppression rate increases for higher energies is consistent with the increasing total cross-section (dashed-line).}
\label{fig4}
\end{figure} 

Figure~\ref{fig1} shows the in-situ resistivity measurement during irradiation. The electron irradiation was performed at $T=$ 22 K in liquid hydrogen. Low temperature is needed to remove the heat generated during irradiation, prevent immediate recombination of Frenkel pairs and, importantly, prevent clusterization and agglomeration of the produced defects. The first irradiation (run 1) with 2.5 MeV electron beam was conducted up to 0.87 C/cm$^2$. During this irradiation, the resistivity monotonically increased from 15 to 30 $\mu \Omega cm$. The rate of resistivity increase per fluence ($\Delta \rho$/$\Delta fluence$) is 16.56 $\mu \Omega cm^{3}/C$. After run 1, the sample is removed from the irradiation chamber and transferred to the other cryostat to measure the temperature dependent resistivity for Fig.~\ref{fig3} (a). For the second irradiation (run 2), the sample is mounted again to the irradiation chamber. Between run1 and run2, the sample is exposed to the room temperature and annealing of defects at room-temperature is evident as a decrease of resistivity from 30 to 24 $\mu \Omega cm$. The run2 was performed with 1.0 MeV electron beam up to 0.21 C/cm$^2$. The identical procedure was repeated for all five irradiation runs in order. 

Figure~\ref{fig2} (a) summarizes the energy dependence of in-situ resistivity found in Fig.~\ref{fig1}. Interestingly, we found that the rate of change of in-situ resistivity, $\Delta \rho / \Delta fluence$, is substantially larger for the irradiation at lower energies. To understand this behavior, we need to calculate the energy-dependent partial cross-section for Ba, Fe, and As. This requires the knowledge of the knock-out barriers, $E_{d}$, which depend on the element and its position in a particular crystal lattice. The knockout threshold barriers $E_{d}$ values, Ba (33 eV), Fe (22 eV) and As (50 eV), were estimated by using Projector Augmented Wave \cite{Blochl_PAW} as implemented in Vienna Ab-initio Simulation Package (VASP) \cite{VASP_ref2}. Gradient correction \cite{perdew1996generalized} was used in the calculations, and semicore Ba-s and Fe-p states were treated as valence states. We used a supercell of 18 formula units and 1 k-point in the Brillouin zone. Ab initio molecular dynamics (MD) was performed using the standard VASP settings \cite{VASP_ref}. Calculations were initialized by assigning a prescribed kinetic energy to a given atom and monitoring whether it will drift away in the process of MD, or return back to its original site. The magnetic state of the starting configuration did not affect the final estimate of the knockout energy within the accuracy that we were interested in. With the obtained $E_{d}$ values, we used SECTE (``Sections Efficaces
Calcul Transport d'\'{E}lectrons") software, developed at \'{E}cole Polytechnique
(Palaiseau, France) by members of the ``Laboratoire des Solides Irradi\'{e}s",
specifically for the interpretation of MeV-range electron irradiation.
Essentially, this is a computer-assisted atomic-weights-averaged interpolation
of the ion knock-out cross-sections tabulated by O. S. Oen \cite{Oen1973}. It appears that the defects produced roughly below 1.5 MeV contribute most to the resistivity change and, according to our calculations, these are defects in the iron sublattice. This is our central profound result, which has always been assumed in iron-based superconductors, but is now verified directly experimentally.

As a next step, we look at the independent parameter that depends on disorder - the superconducting transition temperature, $T_c$.
Figure~\ref{fig3} (a) shows the temperature-dependent resistivity measurement after each irradiation run. The first measurement (pristine) was conducted before irradiation. It has $T_{c, onset}$ = 20.2 K and $T_{c,offset}$ = 19.3 K. After each irradiation, the normal state resistivity increased, indicating the addition of defects. We used the normal state resistivity at 19.5 K, just above the transition, to characterize impurity scattering. Since the $T_c$ of the pristine samples is higher than 19.5 K, we used an extrapolation of the normal state resistivity down 19.5 K to estimate the normal state resistivity. Fig.~\ref{fig3} (b) shows the suppression of $T_{c,onset}$ and $T_{c,offset}$ plotted against the normal state resistivity at $T$ = 19.5 K. In general, $T_{c}$ decreases at the rate of - 0.20 K/$\mu \Omega cm$ ($T_{c,onset}$) and -0.21 K/$\mu \Omega cm$ ($T_{c,offset}$). As expected, $T_c$ is affected by the total increase of resistivity, i.e., the total scattering rate. Defects in all ion sub-lattices contribute to scattering and therefore we should expect that the rate of $T_c$ suppression would depend on the total cross-section. 

The $T_c$ suppression is further analyzed in Fig.~\ref{fig4}. The inset of Fig.~\ref{fig4} explains the way the normalized suppression was calculated during the fourth irradiation with 2.0 MeV and 0.31 C/cm$^2$ (`run4') as an example. $\Delta T_c$ is the variation of $T_c$ before and after 2.0 MeV irradiation, and $\Delta \rho$ is the variation of the resistivity measured at $T$ = 19.5 K before and after 2.0 MeV irradiation. From these values, we calculated a normalized $T_c$ suppression rate of $|\Delta T_c / \Delta \rho|$. The same calculation is performed for all five irradiations and the results are plotted in the main panel of Fig.~\ref{fig4}. Indeed, the normalized $T_c$ suppression rate increases with increasing energy. As asserted above, this is expected since the total cross-section (dashed line) increases with energy. 

\section{Conclusions}
Low-temperature variable-energy electron irradiation was used to probe ion-specific scattering and superconductivity in a single crystal Ba$_{1-x}$K$_x$Fe$_2$As$_2$, $x=$0.80. Measured in-situ at 22 K, the rate of the resistivity increase peaks at electrons energies below 1.5 MeV. Comparison with the calculated partial cross-sections points to the predominant creation of defects in the iron sublattice at these energies. Simultaneously, superconducting $T_c$, measured separately between the irradiation runs, is monotonically suppressed with resistivity increase. This observation reflects that total  scattering rate on all defects, hence total cross-section, monotonically increases with energy. Our work confirms experimentally an often-made assumption of the dominant role of the iron sub-lattice in scattering in iron-based superconductors.

%acknowledgments
\section{Acknowledgements}
This work was supported by the U.S. Department of Energy (DOE), Office of Science, Basic Energy Sciences, Materials Science and Engineering Division. Ames Laboratory is operated for the U.S. DOE by Iowa State University under contract DE-AC02-07CH11358. We thank the SIRIUS team for running electron irradiation at \'{E}cole Polytechnique supported by the EMIR\&A (the French Federation of Accelerators for Irradiation and Analysis of Molecules and Materials) user proposals 15-5788, 17-3646 and 18-5155.

\bibliographystyle{apsrev4-2}
\bibliography{varenergy}

\begin{thebibliography}{38}%
\makeatletter
\providecommand \@ifxundefined [1]{%
 \@ifx{#1\undefined}
}%
\providecommand \@ifnum [1]{%
 \ifnum #1\expandafter \@firstoftwo
 \else \expandafter \@secondoftwo
 \fi
}%
\providecommand \@ifx [1]{%
 \ifx #1\expandafter \@firstoftwo
 \else \expandafter \@secondoftwo
 \fi
}%
\providecommand \natexlab [1]{#1}%
\providecommand \enquote  [1]{``#1''}%
\providecommand \bibnamefont  [1]{#1}%
\providecommand \bibfnamefont [1]{#1}%
\providecommand \citenamefont [1]{#1}%
\providecommand \href@noop [0]{\@secondoftwo}%
\providecommand \href [0]{\begingroup \@sanitize@url \@href}%
\providecommand \@href[1]{\@@startlink{#1}\@@href}%
\providecommand \@@href[1]{\endgroup#1\@@endlink}%
\providecommand \@sanitize@url [0]{\catcode `\\12\catcode `\$12\catcode
  `\&12\catcode `\#12\catcode `\^12\catcode `\_12\catcode `\%12\relax}%
\providecommand \@@startlink[1]{}%
\providecommand \@@endlink[0]{}%
\providecommand \url  [0]{\begingroup\@sanitize@url \@url }%
\providecommand \@url [1]{\endgroup\@href {#1}{\urlprefix }}%
\providecommand \urlprefix  [0]{URL }%
\providecommand \Eprint [0]{\href }%
\providecommand \doibase [0]{https://doi.org/}%
\providecommand \selectlanguage [0]{\@gobble}%
\providecommand \bibinfo  [0]{\@secondoftwo}%
\providecommand \bibfield  [0]{\@secondoftwo}%
\providecommand \translation [1]{[#1]}%
\providecommand \BibitemOpen [0]{}%
\providecommand \bibitemStop [0]{}%
\providecommand \bibitemNoStop [0]{.\EOS\space}%
\providecommand \EOS [0]{\spacefactor3000\relax}%
\providecommand \BibitemShut  [1]{\csname bibitem#1\endcsname}%
\let\auto@bib@innerbib\@empty
%</preamble>
\bibitem [{\citenamefont {Balatsky}\ \emph {et~al.}(2006)\citenamefont
  {Balatsky}, \citenamefont {Vekhter},\ and\ \citenamefont
  {Zhu}}]{Balatsky2006RMP}%
  \BibitemOpen
  \bibfield  {author} {\bibinfo {author} {\bibfnamefont {A.~V.}\ \bibnamefont
  {Balatsky}}, \bibinfo {author} {\bibfnamefont {I.}~\bibnamefont {Vekhter}},\
  and\ \bibinfo {author} {\bibfnamefont {J.-X.}\ \bibnamefont {Zhu}},\ }\href
  {https://doi.org/10.1103/RevModPhys.78.373} {\bibfield  {journal} {\bibinfo
  {journal} {Rev. Mod. Phys.}\ }\textbf {\bibinfo {volume} {78}},\ \bibinfo
  {pages} {373} (\bibinfo {year} {2006})}\BibitemShut {NoStop}%
\bibitem [{\citenamefont {Alloul}\ \emph {et~al.}(2009)\citenamefont {Alloul},
  \citenamefont {Bobroff}, \citenamefont {Gabay},\ and\ \citenamefont
  {Hirschfeld}}]{Alloul2009RMP_defect_cuprates}%
  \BibitemOpen
  \bibfield  {author} {\bibinfo {author} {\bibfnamefont {H.}~\bibnamefont
  {Alloul}}, \bibinfo {author} {\bibfnamefont {J.}~\bibnamefont {Bobroff}},
  \bibinfo {author} {\bibfnamefont {M.}~\bibnamefont {Gabay}},\ and\ \bibinfo
  {author} {\bibfnamefont {P.~J.}\ \bibnamefont {Hirschfeld}},\ }\href
  {https://doi.org/10.1103/RevModPhys.81.45} {\bibfield  {journal} {\bibinfo
  {journal} {Rev. Mod. Phys.}\ }\textbf {\bibinfo {volume} {81}},\ \bibinfo
  {pages} {45} (\bibinfo {year} {2009})}\BibitemShut {NoStop}%
\bibitem [{\citenamefont {Li}\ \emph {et~al.}(2016)\citenamefont {Li},
  \citenamefont {Guo}, \citenamefont {Yang}, \citenamefont {Yamaura},
  \citenamefont {Takayama-Muromachi}, \citenamefont {Wang},\ and\ \citenamefont
  {Wu}}]{LiWu2016SST_review_FeSC_chem_impurity}%
  \BibitemOpen
  \bibfield  {author} {\bibinfo {author} {\bibfnamefont {J.}~\bibnamefont
  {Li}}, \bibinfo {author} {\bibfnamefont {Y.-F.}\ \bibnamefont {Guo}},
  \bibinfo {author} {\bibfnamefont {Z.-R.}\ \bibnamefont {Yang}}, \bibinfo
  {author} {\bibfnamefont {K.}~\bibnamefont {Yamaura}}, \bibinfo {author}
  {\bibfnamefont {E.}~\bibnamefont {Takayama-Muromachi}}, \bibinfo {author}
  {\bibfnamefont {H.-B.}\ \bibnamefont {Wang}},\ and\ \bibinfo {author}
  {\bibfnamefont {P.-H.}\ \bibnamefont {Wu}},\ }\href
  {https://doi.org/10.1088/0953-2048/29/5/053001} {\bibfield  {journal}
  {\bibinfo  {journal} {Superconductor Science and Technology}\ }\textbf
  {\bibinfo {volume} {29}},\ \bibinfo {pages} {053001} (\bibinfo {year}
  {2016})}\BibitemShut {NoStop}%
\bibitem [{\citenamefont {Anderson}(1959)}]{Anderson1959JPCS}%
  \BibitemOpen
  \bibfield  {author} {\bibinfo {author} {\bibfnamefont {P.}~\bibnamefont
  {Anderson}},\ }\href
  {https://doi.org/https://doi.org/10.1016/0022-3697(59)90036-8} {\bibfield
  {journal} {\bibinfo  {journal} {Journal of Physics and Chemistry of Solids}\
  }\textbf {\bibinfo {volume} {11}},\ \bibinfo {pages} {26} (\bibinfo {year}
  {1959})}\BibitemShut {NoStop}%
\bibitem [{\citenamefont {Abrikosov}\ and\ \citenamefont
  {Gor'kov}(1961)}]{AbrikosovGorkov1961JETP}%
  \BibitemOpen
  \bibfield  {author} {\bibinfo {author} {\bibfnamefont {A.~A.}\ \bibnamefont
  {Abrikosov}}\ and\ \bibinfo {author} {\bibfnamefont {L.~P.}\ \bibnamefont
  {Gor'kov}},\ }\href@noop {} {\bibfield  {journal} {\bibinfo  {journal}
  {Soviet Physics JETP-USSR}\ }\textbf {\bibinfo {volume} {12}},\ \bibinfo
  {pages} {1243} (\bibinfo {year} {1961})}\BibitemShut {NoStop}%
\bibitem [{\citenamefont {Openov}(1998)}]{Openov1998PRB_impurity}%
  \BibitemOpen
  \bibfield  {author} {\bibinfo {author} {\bibfnamefont {L.~A.}\ \bibnamefont
  {Openov}},\ }\href {https://doi.org/10.1103/PhysRevB.58.9468} {\bibfield
  {journal} {\bibinfo  {journal} {Phys. Rev. B}\ }\textbf {\bibinfo {volume}
  {58}},\ \bibinfo {pages} {9468} (\bibinfo {year} {1998})}\BibitemShut
  {NoStop}%
\bibitem [{\citenamefont {Kogan}\ \emph {et~al.}(2013)\citenamefont {Kogan},
  \citenamefont {Prozorov},\ and\ \citenamefont
  {Mishra}}]{KoganProzorovMishra2013PRB}%
  \BibitemOpen
  \bibfield  {author} {\bibinfo {author} {\bibfnamefont {V.~G.}\ \bibnamefont
  {Kogan}}, \bibinfo {author} {\bibfnamefont {R.}~\bibnamefont {Prozorov}},\
  and\ \bibinfo {author} {\bibfnamefont {V.}~\bibnamefont {Mishra}},\ }\href
  {https://doi.org/10.1103/PhysRevB.88.224508} {\bibfield  {journal} {\bibinfo
  {journal} {Phys. Rev. B}\ }\textbf {\bibinfo {volume} {88}},\ \bibinfo
  {pages} {224508} (\bibinfo {year} {2013})}\BibitemShut {NoStop}%
\bibitem [{\citenamefont {Finnemore}\ \emph {et~al.}(1965)\citenamefont
  {Finnemore}, \citenamefont {Johnson}, \citenamefont {Ostenson}, \citenamefont
  {Spedding},\ and\ \citenamefont {Beaudry}}]{Finnemore1965PR}%
  \BibitemOpen
  \bibfield  {author} {\bibinfo {author} {\bibfnamefont {D.~K.}\ \bibnamefont
  {Finnemore}}, \bibinfo {author} {\bibfnamefont {D.~L.}\ \bibnamefont
  {Johnson}}, \bibinfo {author} {\bibfnamefont {J.~E.}\ \bibnamefont
  {Ostenson}}, \bibinfo {author} {\bibfnamefont {F.~H.}\ \bibnamefont
  {Spedding}},\ and\ \bibinfo {author} {\bibfnamefont {B.~J.}\ \bibnamefont
  {Beaudry}},\ }\href {https://doi.org/10.1103/PhysRev.137.A550} {\bibfield
  {journal} {\bibinfo  {journal} {Phys. Rev.}\ }\textbf {\bibinfo {volume}
  {137}},\ \bibinfo {pages} {A550} (\bibinfo {year} {1965})}\BibitemShut
  {NoStop}%
\bibitem [{\citenamefont {Giapintzakis}\ \emph {et~al.}(1994)\citenamefont
  {Giapintzakis}, \citenamefont {Ginsberg}, \citenamefont {Kirk},\ and\
  \citenamefont {Ockers}}]{Giapintzakis1994PRB_YBCO_e-irr}%
  \BibitemOpen
  \bibfield  {author} {\bibinfo {author} {\bibfnamefont {J.}~\bibnamefont
  {Giapintzakis}}, \bibinfo {author} {\bibfnamefont {D.~M.}\ \bibnamefont
  {Ginsberg}}, \bibinfo {author} {\bibfnamefont {M.~A.}\ \bibnamefont {Kirk}},\
  and\ \bibinfo {author} {\bibfnamefont {S.}~\bibnamefont {Ockers}},\ }\href
  {https://doi.org/10.1103/PhysRevB.50.15967} {\bibfield  {journal} {\bibinfo
  {journal} {Phys. Rev. B}\ }\textbf {\bibinfo {volume} {50}},\ \bibinfo
  {pages} {15967} (\bibinfo {year} {1994})}\BibitemShut {NoStop}%
\bibitem [{\citenamefont {Cho}\ \emph {et~al.}(2016)\citenamefont {Cho},
  \citenamefont {Ko\'{n}czykowski}, \citenamefont {Teknowijoyo}, \citenamefont
  {Tanatar}, \citenamefont {Liu}, \citenamefont {Lograsso}, \citenamefont
  {Straszheim}, \citenamefont {Mishra}, \citenamefont {Maiti}, \citenamefont
  {Hirschfeld},\ and\ \citenamefont
  {Prozorov}}]{Cho2016ScienceAdvances_BaK122_e-irr}%
  \BibitemOpen
  \bibfield  {author} {\bibinfo {author} {\bibfnamefont {K.}~\bibnamefont
  {Cho}}, \bibinfo {author} {\bibfnamefont {M.}~\bibnamefont
  {Ko\'{n}czykowski}}, \bibinfo {author} {\bibfnamefont {S.}~\bibnamefont
  {Teknowijoyo}}, \bibinfo {author} {\bibfnamefont {M.~A.}\ \bibnamefont
  {Tanatar}}, \bibinfo {author} {\bibfnamefont {Y.}~\bibnamefont {Liu}},
  \bibinfo {author} {\bibfnamefont {T.~A.}\ \bibnamefont {Lograsso}}, \bibinfo
  {author} {\bibfnamefont {W.~E.}\ \bibnamefont {Straszheim}}, \bibinfo
  {author} {\bibfnamefont {V.}~\bibnamefont {Mishra}}, \bibinfo {author}
  {\bibfnamefont {S.}~\bibnamefont {Maiti}}, \bibinfo {author} {\bibfnamefont
  {P.~J.}\ \bibnamefont {Hirschfeld}},\ and\ \bibinfo {author} {\bibfnamefont
  {R.}~\bibnamefont {Prozorov}},\ }\href
  {https://doi.org/10.1126/sciadv.1600807} {\bibfield  {journal} {\bibinfo
  {journal} {Science Advances}\ }\textbf {\bibinfo {volume} {2}},\ \bibinfo
  {pages} {e1600807} (\bibinfo {year} {2016})}\BibitemShut {NoStop}%
\bibitem [{\citenamefont {Cho}\ \emph {et~al.}(2018{\natexlab{a}})\citenamefont
  {Cho}, \citenamefont {Ko{\'n}czykowski}, \citenamefont {Teknowijoyo},
  \citenamefont {Tanatar},\ and\ \citenamefont
  {Prozorov}}]{Cho2018SST_review_e-irr}%
  \BibitemOpen
  \bibfield  {author} {\bibinfo {author} {\bibfnamefont {K.}~\bibnamefont
  {Cho}}, \bibinfo {author} {\bibfnamefont {M.}~\bibnamefont
  {Ko{\'n}czykowski}}, \bibinfo {author} {\bibfnamefont {S.}~\bibnamefont
  {Teknowijoyo}}, \bibinfo {author} {\bibfnamefont {M.~A.}\ \bibnamefont
  {Tanatar}},\ and\ \bibinfo {author} {\bibfnamefont {R.}~\bibnamefont
  {Prozorov}},\ }\href@noop {} {\bibfield  {journal} {\bibinfo  {journal}
  {Superconductor Science and Technology}\ }\textbf {\bibinfo {volume} {31}},\
  \bibinfo {pages} {064002} (\bibinfo {year} {2018}{\natexlab{a}})}\BibitemShut
  {NoStop}%
\bibitem [{\citenamefont {Cho}\ \emph {et~al.}(2018{\natexlab{b}})\citenamefont
  {Cho}, \citenamefont {Ko{\'{n}}czykowski}, \citenamefont {Teknowijoyo},
  \citenamefont {Tanatar}, \citenamefont {Guss}, \citenamefont {Gartin},
  \citenamefont {Wilde}, \citenamefont {Kreyssig}, \citenamefont {McQueeney},
  \citenamefont {Goldman}, \citenamefont {Mishra}, \citenamefont {Hirschfeld},\
  and\ \citenamefont {Prozorov}}]{Cho2018NatComm_NbSe2}%
  \BibitemOpen
  \bibfield  {author} {\bibinfo {author} {\bibfnamefont {K.}~\bibnamefont
  {Cho}}, \bibinfo {author} {\bibfnamefont {M.}~\bibnamefont
  {Ko{\'{n}}czykowski}}, \bibinfo {author} {\bibfnamefont {S.}~\bibnamefont
  {Teknowijoyo}}, \bibinfo {author} {\bibfnamefont {M.~A.}\ \bibnamefont
  {Tanatar}}, \bibinfo {author} {\bibfnamefont {J.}~\bibnamefont {Guss}},
  \bibinfo {author} {\bibfnamefont {P.~B.}\ \bibnamefont {Gartin}}, \bibinfo
  {author} {\bibfnamefont {J.~M.}\ \bibnamefont {Wilde}}, \bibinfo {author}
  {\bibfnamefont {A.}~\bibnamefont {Kreyssig}}, \bibinfo {author}
  {\bibfnamefont {R.~J.}\ \bibnamefont {McQueeney}}, \bibinfo {author}
  {\bibfnamefont {A.~I.}\ \bibnamefont {Goldman}}, \bibinfo {author}
  {\bibfnamefont {V.}~\bibnamefont {Mishra}}, \bibinfo {author} {\bibfnamefont
  {P.~J.}\ \bibnamefont {Hirschfeld}},\ and\ \bibinfo {author} {\bibfnamefont
  {R.}~\bibnamefont {Prozorov}},\ }\href
  {https://doi.org/10.1038/s41467-018-05153-0} {\bibfield  {journal} {\bibinfo
  {journal} {Nature Communications}\ }\textbf {\bibinfo {volume} {9}},\
  \bibinfo {pages} {2796} (\bibinfo {year} {2018}{\natexlab{b}})}\BibitemShut
  {NoStop}%
\bibitem [{\citenamefont {Cho}\ \emph {et~al.}(2022)\citenamefont {Cho},
  \citenamefont {Ko\ifmmode~\acute{n}\else \'{n}\fi{}czykowski}, \citenamefont
  {Teknowijoyo}, \citenamefont {Ghimire}, \citenamefont {Tanatar},
  \citenamefont {Mishra},\ and\ \citenamefont
  {Prozorov}}]{Cho2022PRB_YBCO_e-irr}%
  \BibitemOpen
  \bibfield  {author} {\bibinfo {author} {\bibfnamefont {K.}~\bibnamefont
  {Cho}}, \bibinfo {author} {\bibfnamefont {M.}~\bibnamefont
  {Ko\ifmmode~\acute{n}\else \'{n}\fi{}czykowski}}, \bibinfo {author}
  {\bibfnamefont {S.}~\bibnamefont {Teknowijoyo}}, \bibinfo {author}
  {\bibfnamefont {S.}~\bibnamefont {Ghimire}}, \bibinfo {author} {\bibfnamefont
  {M.~A.}\ \bibnamefont {Tanatar}}, \bibinfo {author} {\bibfnamefont
  {V.}~\bibnamefont {Mishra}},\ and\ \bibinfo {author} {\bibfnamefont
  {R.}~\bibnamefont {Prozorov}},\ }\href
  {https://doi.org/10.1103/PhysRevB.105.014514} {\bibfield  {journal} {\bibinfo
   {journal} {Phys. Rev. B}\ }\textbf {\bibinfo {volume} {105}},\ \bibinfo
  {pages} {014514} (\bibinfo {year} {2022})}\BibitemShut {NoStop}%
\bibitem [{\citenamefont {Nakajima}\ \emph {et~al.}(2010)\citenamefont
  {Nakajima}, \citenamefont {Taen}, \citenamefont {Tsuchiya}, \citenamefont
  {Tamegai}, \citenamefont {Kitamura},\ and\ \citenamefont
  {Murakami}}]{Nakajima2010PRB_BaCo122_proton_irr}%
  \BibitemOpen
  \bibfield  {author} {\bibinfo {author} {\bibfnamefont {Y.}~\bibnamefont
  {Nakajima}}, \bibinfo {author} {\bibfnamefont {T.}~\bibnamefont {Taen}},
  \bibinfo {author} {\bibfnamefont {Y.}~\bibnamefont {Tsuchiya}}, \bibinfo
  {author} {\bibfnamefont {T.}~\bibnamefont {Tamegai}}, \bibinfo {author}
  {\bibfnamefont {H.}~\bibnamefont {Kitamura}},\ and\ \bibinfo {author}
  {\bibfnamefont {T.}~\bibnamefont {Murakami}},\ }\href
  {https://doi.org/10.1103/PhysRevB.82.220504} {\bibfield  {journal} {\bibinfo
  {journal} {Phys. Rev. B}\ }\textbf {\bibinfo {volume} {82}},\ \bibinfo
  {pages} {220504} (\bibinfo {year} {2010})}\BibitemShut {NoStop}%
\bibitem [{\citenamefont {Taen}\ \emph {et~al.}(2013)\citenamefont {Taen},
  \citenamefont {Ohtake}, \citenamefont {Akiyama}, \citenamefont {Inoue},
  \citenamefont {Sun}, \citenamefont {Pyon}, \citenamefont {Tamegai},\ and\
  \citenamefont {Kitamura}}]{Taen2013PRB_BaK122_proton_irr}%
  \BibitemOpen
  \bibfield  {author} {\bibinfo {author} {\bibfnamefont {T.}~\bibnamefont
  {Taen}}, \bibinfo {author} {\bibfnamefont {F.}~\bibnamefont {Ohtake}},
  \bibinfo {author} {\bibfnamefont {H.}~\bibnamefont {Akiyama}}, \bibinfo
  {author} {\bibfnamefont {H.}~\bibnamefont {Inoue}}, \bibinfo {author}
  {\bibfnamefont {Y.}~\bibnamefont {Sun}}, \bibinfo {author} {\bibfnamefont
  {S.}~\bibnamefont {Pyon}}, \bibinfo {author} {\bibfnamefont {T.}~\bibnamefont
  {Tamegai}},\ and\ \bibinfo {author} {\bibfnamefont {H.}~\bibnamefont
  {Kitamura}},\ }\href {https://doi.org/10.1103/PhysRevB.88.224514} {\bibfield
  {journal} {\bibinfo  {journal} {Phys. Rev. B}\ }\textbf {\bibinfo {volume}
  {88}},\ \bibinfo {pages} {224514} (\bibinfo {year} {2013})}\BibitemShut
  {NoStop}%
\bibitem [{\citenamefont {Smylie}\ \emph {et~al.}(2016)\citenamefont {Smylie},
  \citenamefont {Leroux}, \citenamefont {Mishra}, \citenamefont {Fang},
  \citenamefont {Taddei}, \citenamefont {Chmaissem}, \citenamefont {Claus},
  \citenamefont {Kayani}, \citenamefont {Snezhko}, \citenamefont {Welp},\ and\
  \citenamefont {Kwok}}]{Smylie2016PRB_BaP122_proton_irr}%
  \BibitemOpen
  \bibfield  {author} {\bibinfo {author} {\bibfnamefont {M.~P.}\ \bibnamefont
  {Smylie}}, \bibinfo {author} {\bibfnamefont {M.}~\bibnamefont {Leroux}},
  \bibinfo {author} {\bibfnamefont {V.}~\bibnamefont {Mishra}}, \bibinfo
  {author} {\bibfnamefont {L.}~\bibnamefont {Fang}}, \bibinfo {author}
  {\bibfnamefont {K.~M.}\ \bibnamefont {Taddei}}, \bibinfo {author}
  {\bibfnamefont {O.}~\bibnamefont {Chmaissem}}, \bibinfo {author}
  {\bibfnamefont {H.}~\bibnamefont {Claus}}, \bibinfo {author} {\bibfnamefont
  {A.}~\bibnamefont {Kayani}}, \bibinfo {author} {\bibfnamefont
  {A.}~\bibnamefont {Snezhko}}, \bibinfo {author} {\bibfnamefont
  {U.}~\bibnamefont {Welp}},\ and\ \bibinfo {author} {\bibfnamefont {W.-K.}\
  \bibnamefont {Kwok}},\ }\href {https://doi.org/10.1103/PhysRevB.93.115119}
  {\bibfield  {journal} {\bibinfo  {journal} {Phys. Rev. B}\ }\textbf {\bibinfo
  {volume} {93}},\ \bibinfo {pages} {115119} (\bibinfo {year}
  {2016})}\BibitemShut {NoStop}%
\bibitem [{\citenamefont {Moroni}\ \emph {et~al.}(2017)\citenamefont {Moroni},
  \citenamefont {Gozzelino}, \citenamefont {Ghigo}, \citenamefont {Tanatar},
  \citenamefont {Prozorov}, \citenamefont {Canfield},\ and\ \citenamefont
  {Carretta}}]{Moroni2017PRB_proton-irr}%
  \BibitemOpen
  \bibfield  {author} {\bibinfo {author} {\bibfnamefont {M.}~\bibnamefont
  {Moroni}}, \bibinfo {author} {\bibfnamefont {L.}~\bibnamefont {Gozzelino}},
  \bibinfo {author} {\bibfnamefont {G.}~\bibnamefont {Ghigo}}, \bibinfo
  {author} {\bibfnamefont {M.~A.}\ \bibnamefont {Tanatar}}, \bibinfo {author}
  {\bibfnamefont {R.}~\bibnamefont {Prozorov}}, \bibinfo {author}
  {\bibfnamefont {P.~C.}\ \bibnamefont {Canfield}},\ and\ \bibinfo {author}
  {\bibfnamefont {P.}~\bibnamefont {Carretta}},\ }\href
  {https://doi.org/10.1103/PhysRevB.96.094523} {\bibfield  {journal} {\bibinfo
  {journal} {Phys. Rev. B}\ }\textbf {\bibinfo {volume} {96}},\ \bibinfo
  {pages} {094523} (\bibinfo {year} {2017})}\BibitemShut {NoStop}%
\bibitem [{\citenamefont {Sylva}\ \emph {et~al.}(2018)\citenamefont {Sylva},
  \citenamefont {Bellingeri}, \citenamefont {Ferdeghini}, \citenamefont
  {Martinelli}, \citenamefont {Pallecchi}, \citenamefont {Pellegrino},
  \citenamefont {Putti}, \citenamefont {Ghigo}, \citenamefont {Gozzelino},
  \citenamefont {Torsello}, \citenamefont {Grimaldi}, \citenamefont {Leo},
  \citenamefont {Nigro},\ and\ \citenamefont
  {Braccini}}]{Sylva2018SST_FeSe_thinfilm_proton-irr}%
  \BibitemOpen
  \bibfield  {author} {\bibinfo {author} {\bibfnamefont {G.}~\bibnamefont
  {Sylva}}, \bibinfo {author} {\bibfnamefont {E.}~\bibnamefont {Bellingeri}},
  \bibinfo {author} {\bibfnamefont {C.}~\bibnamefont {Ferdeghini}}, \bibinfo
  {author} {\bibfnamefont {A.}~\bibnamefont {Martinelli}}, \bibinfo {author}
  {\bibfnamefont {I.}~\bibnamefont {Pallecchi}}, \bibinfo {author}
  {\bibfnamefont {L.}~\bibnamefont {Pellegrino}}, \bibinfo {author}
  {\bibfnamefont {M.}~\bibnamefont {Putti}}, \bibinfo {author} {\bibfnamefont
  {G.}~\bibnamefont {Ghigo}}, \bibinfo {author} {\bibfnamefont
  {L.}~\bibnamefont {Gozzelino}}, \bibinfo {author} {\bibfnamefont
  {D.}~\bibnamefont {Torsello}}, \bibinfo {author} {\bibfnamefont
  {G.}~\bibnamefont {Grimaldi}}, \bibinfo {author} {\bibfnamefont
  {A.}~\bibnamefont {Leo}}, \bibinfo {author} {\bibfnamefont {A.}~\bibnamefont
  {Nigro}},\ and\ \bibinfo {author} {\bibfnamefont {V.}~\bibnamefont
  {Braccini}},\ }\href {https://doi.org/10.1088/1361-6668/aab3bd} {\bibfield
  {journal} {\bibinfo  {journal} {Superconductor Science and Technology}\
  }\textbf {\bibinfo {volume} {31}},\ \bibinfo {pages} {054001} (\bibinfo
  {year} {2018})}\BibitemShut {NoStop}%
\bibitem [{\citenamefont {Torsello}\ \emph {et~al.}(2022)\citenamefont
  {Torsello}, \citenamefont {Fracasso}, \citenamefont {Gerbaldo}, \citenamefont
  {Ghigo}, \citenamefont {Laviano}, \citenamefont {Napolitano}, \citenamefont
  {Iebole}, \citenamefont {Cialone}, \citenamefont {Manca}, \citenamefont
  {Martinelli}, \citenamefont {Piperno}, \citenamefont {Braccini},
  \citenamefont {Leo}, \citenamefont {Grimaldi}, \citenamefont {Vannozzi},
  \citenamefont {Celentano}, \citenamefont {Putti},\ and\ \citenamefont
  {Gozzelino}}]{Torsello2022IEEE_FeSe_thinfilm_proton-irr}%
  \BibitemOpen
  \bibfield  {author} {\bibinfo {author} {\bibfnamefont {D.}~\bibnamefont
  {Torsello}}, \bibinfo {author} {\bibfnamefont {M.}~\bibnamefont {Fracasso}},
  \bibinfo {author} {\bibfnamefont {R.}~\bibnamefont {Gerbaldo}}, \bibinfo
  {author} {\bibfnamefont {G.}~\bibnamefont {Ghigo}}, \bibinfo {author}
  {\bibfnamefont {F.}~\bibnamefont {Laviano}}, \bibinfo {author} {\bibfnamefont
  {A.}~\bibnamefont {Napolitano}}, \bibinfo {author} {\bibfnamefont
  {M.}~\bibnamefont {Iebole}}, \bibinfo {author} {\bibfnamefont
  {M.}~\bibnamefont {Cialone}}, \bibinfo {author} {\bibfnamefont
  {N.}~\bibnamefont {Manca}}, \bibinfo {author} {\bibfnamefont
  {A.}~\bibnamefont {Martinelli}}, \bibinfo {author} {\bibfnamefont
  {L.}~\bibnamefont {Piperno}}, \bibinfo {author} {\bibfnamefont
  {V.}~\bibnamefont {Braccini}}, \bibinfo {author} {\bibfnamefont
  {A.}~\bibnamefont {Leo}}, \bibinfo {author} {\bibfnamefont {G.}~\bibnamefont
  {Grimaldi}}, \bibinfo {author} {\bibfnamefont {A.}~\bibnamefont {Vannozzi}},
  \bibinfo {author} {\bibfnamefont {G.}~\bibnamefont {Celentano}}, \bibinfo
  {author} {\bibfnamefont {M.}~\bibnamefont {Putti}},\ and\ \bibinfo {author}
  {\bibfnamefont {L.}~\bibnamefont {Gozzelino}},\ }\href
  {https://doi.org/10.1109/TASC.2021.3136135} {\bibfield  {journal} {\bibinfo
  {journal} {IEEE Transactions on Applied Superconductivity}\ }\textbf
  {\bibinfo {volume} {32}},\ \bibinfo {pages} {1} (\bibinfo {year}
  {2022})}\BibitemShut {NoStop}%
\bibitem [{\citenamefont {Konczykowski}\ \emph {et~al.}(1991)\citenamefont
  {Konczykowski}, \citenamefont {Rullier-Albenque}, \citenamefont {Yacoby},
  \citenamefont {Shaulov}, \citenamefont {Yeshurun},\ and\ \citenamefont
  {Lejay}}]{KonczykowskiPRB1991_HeavyIon_YBCO}%
  \BibitemOpen
  \bibfield  {author} {\bibinfo {author} {\bibfnamefont {M.}~\bibnamefont
  {Konczykowski}}, \bibinfo {author} {\bibfnamefont {F.}~\bibnamefont
  {Rullier-Albenque}}, \bibinfo {author} {\bibfnamefont {E.~R.}\ \bibnamefont
  {Yacoby}}, \bibinfo {author} {\bibfnamefont {A.}~\bibnamefont {Shaulov}},
  \bibinfo {author} {\bibfnamefont {Y.}~\bibnamefont {Yeshurun}},\ and\
  \bibinfo {author} {\bibfnamefont {P.}~\bibnamefont {Lejay}},\ }\href
  {https://doi.org/10.1103/PhysRevB.44.7167} {\bibfield  {journal} {\bibinfo
  {journal} {Phys. Rev. B}\ }\textbf {\bibinfo {volume} {44}},\ \bibinfo
  {pages} {7167} (\bibinfo {year} {1991})}\BibitemShut {NoStop}%
\bibitem [{\citenamefont {Nakajima}\ \emph {et~al.}(2009)\citenamefont
  {Nakajima}, \citenamefont {Tsuchiya}, \citenamefont {Taen}, \citenamefont
  {Tamegai}, \citenamefont {Okayasu},\ and\ \citenamefont
  {Sasase}}]{Nakajima2009PRB_HeavyIon}%
  \BibitemOpen
  \bibfield  {author} {\bibinfo {author} {\bibfnamefont {Y.}~\bibnamefont
  {Nakajima}}, \bibinfo {author} {\bibfnamefont {Y.}~\bibnamefont {Tsuchiya}},
  \bibinfo {author} {\bibfnamefont {T.}~\bibnamefont {Taen}}, \bibinfo {author}
  {\bibfnamefont {T.}~\bibnamefont {Tamegai}}, \bibinfo {author} {\bibfnamefont
  {S.}~\bibnamefont {Okayasu}},\ and\ \bibinfo {author} {\bibfnamefont
  {M.}~\bibnamefont {Sasase}},\ }\href
  {https://doi.org/10.1103/PhysRevB.80.012510} {\bibfield  {journal} {\bibinfo
  {journal} {Phys. Rev. B}\ }\textbf {\bibinfo {volume} {80}},\ \bibinfo
  {pages} {012510} (\bibinfo {year} {2009})}\BibitemShut {NoStop}%
\bibitem [{\citenamefont {Prozorov}\ \emph {et~al.}(2010)\citenamefont
  {Prozorov}, \citenamefont {Tanatar}, \citenamefont {Roy}, \citenamefont {Ni},
  \citenamefont {Bud'ko}, \citenamefont {Canfield}, \citenamefont {Hua},
  \citenamefont {Welp},\ and\ \citenamefont {Kwok}}]{Prozorov2010PRB_HeavyIon}%
  \BibitemOpen
  \bibfield  {author} {\bibinfo {author} {\bibfnamefont {R.}~\bibnamefont
  {Prozorov}}, \bibinfo {author} {\bibfnamefont {M.~A.}\ \bibnamefont
  {Tanatar}}, \bibinfo {author} {\bibfnamefont {B.}~\bibnamefont {Roy}},
  \bibinfo {author} {\bibfnamefont {N.}~\bibnamefont {Ni}}, \bibinfo {author}
  {\bibfnamefont {S.~L.}\ \bibnamefont {Bud'ko}}, \bibinfo {author}
  {\bibfnamefont {P.~C.}\ \bibnamefont {Canfield}}, \bibinfo {author}
  {\bibfnamefont {J.}~\bibnamefont {Hua}}, \bibinfo {author} {\bibfnamefont
  {U.}~\bibnamefont {Welp}},\ and\ \bibinfo {author} {\bibfnamefont {W.~K.}\
  \bibnamefont {Kwok}},\ }\href {https://doi.org/10.1103/PhysRevB.81.094509}
  {\bibfield  {journal} {\bibinfo  {journal} {Phys. Rev. B}\ }\textbf {\bibinfo
  {volume} {81}},\ \bibinfo {pages} {094509} (\bibinfo {year}
  {2010})}\BibitemShut {NoStop}%
\bibitem [{\citenamefont {Murphy}\ \emph {et~al.}(2013)\citenamefont {Murphy},
  \citenamefont {Tanatar}, \citenamefont {Kim}, \citenamefont {Kwok},
  \citenamefont {Welp}, \citenamefont {Graf}, \citenamefont {Brooks},
  \citenamefont {Bud'ko}, \citenamefont {Canfield},\ and\ \citenamefont
  {Prozorov}}]{MurphyProzorov2013PRB_heavy_ion}%
  \BibitemOpen
  \bibfield  {author} {\bibinfo {author} {\bibfnamefont {J.}~\bibnamefont
  {Murphy}}, \bibinfo {author} {\bibfnamefont {M.~A.}\ \bibnamefont {Tanatar}},
  \bibinfo {author} {\bibfnamefont {H.}~\bibnamefont {Kim}}, \bibinfo {author}
  {\bibfnamefont {W.}~\bibnamefont {Kwok}}, \bibinfo {author} {\bibfnamefont
  {U.}~\bibnamefont {Welp}}, \bibinfo {author} {\bibfnamefont {D.}~\bibnamefont
  {Graf}}, \bibinfo {author} {\bibfnamefont {J.~S.}\ \bibnamefont {Brooks}},
  \bibinfo {author} {\bibfnamefont {S.~L.}\ \bibnamefont {Bud'ko}}, \bibinfo
  {author} {\bibfnamefont {P.~C.}\ \bibnamefont {Canfield}},\ and\ \bibinfo
  {author} {\bibfnamefont {R.}~\bibnamefont {Prozorov}},\ }\href
  {https://doi.org/10.1103/PhysRevB.88.054514} {\bibfield  {journal} {\bibinfo
  {journal} {Phys. Rev. B}\ }\textbf {\bibinfo {volume} {88}},\ \bibinfo
  {pages} {054514} (\bibinfo {year} {2013})}\BibitemShut {NoStop}%
\bibitem [{\citenamefont {Damask}\ and\ \citenamefont
  {Dienes}(1963)}]{Damask1963PointDefectsInMetals}%
  \BibitemOpen
  \bibfield  {author} {\bibinfo {author} {\bibfnamefont {A.~C.}\ \bibnamefont
  {Damask}}\ and\ \bibinfo {author} {\bibfnamefont {G.~J.}\ \bibnamefont
  {Dienes}},\ }\href@noop {} {\emph {\bibinfo {title} {Point Defects in
  Metals}}}\ (\bibinfo  {publisher} {Gordon \& Breach Science Publishers Ltd,
  London},\ \bibinfo {year} {1963})\BibitemShut {NoStop}%
\bibitem [{\citenamefont {Thompson}(1969)}]{Thompson1969}%
  \BibitemOpen
  \bibfield  {author} {\bibinfo {author} {\bibfnamefont {M.~W.}\ \bibnamefont
  {Thompson}},\ }\href@noop {} {\emph {\bibinfo {title} {{D}efects and
  {R}adiation {D}amage in {M}etals}}},\ \bibinfo {edition} {revised september
  27, 1974}\ ed.,\ Cambridge Monographs on Physics\ (\bibinfo  {publisher}
  {Cambridge University Press},\ \bibinfo {year} {1969})\BibitemShut {NoStop}%
\bibitem [{\citenamefont {Dai}(2015)}]{Dai2015RMP_revew_FeSCs_Neutron}%
  \BibitemOpen
  \bibfield  {author} {\bibinfo {author} {\bibfnamefont {P.}~\bibnamefont
  {Dai}},\ }\href {https://doi.org/10.1103/RevModPhys.87.855} {\bibfield
  {journal} {\bibinfo  {journal} {Rev. Mod. Phys.}\ }\textbf {\bibinfo {volume}
  {87}},\ \bibinfo {pages} {855} (\bibinfo {year} {2015})}\BibitemShut
  {NoStop}%
\bibitem [{\citenamefont {Fernandes}\ and\ \citenamefont
  {Chubukov}(2016)}]{Fernandes2017RPP_review}%
  \BibitemOpen
  \bibfield  {author} {\bibinfo {author} {\bibfnamefont {R.~M.}\ \bibnamefont
  {Fernandes}}\ and\ \bibinfo {author} {\bibfnamefont {A.~V.}\ \bibnamefont
  {Chubukov}},\ }\href {https://doi.org/10.1088/1361-6633/80/1/014503}
  {\bibfield  {journal} {\bibinfo  {journal} {Reports on Progress in Physics}\
  }\textbf {\bibinfo {volume} {80}},\ \bibinfo {pages} {014503} (\bibinfo
  {year} {2016})}\BibitemShut {NoStop}%
\bibitem [{\citenamefont {Xu}\ \emph {et~al.}(2013)\citenamefont {Xu},
  \citenamefont {Richard}, \citenamefont {Shi}, \citenamefont {van Roekeghem},
  \citenamefont {Qian}, \citenamefont {Razzoli}, \citenamefont {Rienks},
  \citenamefont {Chen}, \citenamefont {Ieki}, \citenamefont {Nakayama},
  \citenamefont {Sato}, \citenamefont {Takahashi}, \citenamefont {Shi},\ and\
  \citenamefont {Ding}}]{Xu2013PRB_BaK122}%
  \BibitemOpen
  \bibfield  {author} {\bibinfo {author} {\bibfnamefont {N.}~\bibnamefont
  {Xu}}, \bibinfo {author} {\bibfnamefont {P.}~\bibnamefont {Richard}},
  \bibinfo {author} {\bibfnamefont {X.}~\bibnamefont {Shi}}, \bibinfo {author}
  {\bibfnamefont {A.}~\bibnamefont {van Roekeghem}}, \bibinfo {author}
  {\bibfnamefont {T.}~\bibnamefont {Qian}}, \bibinfo {author} {\bibfnamefont
  {E.}~\bibnamefont {Razzoli}}, \bibinfo {author} {\bibfnamefont
  {E.}~\bibnamefont {Rienks}}, \bibinfo {author} {\bibfnamefont {G.-F.}\
  \bibnamefont {Chen}}, \bibinfo {author} {\bibfnamefont {E.}~\bibnamefont
  {Ieki}}, \bibinfo {author} {\bibfnamefont {K.}~\bibnamefont {Nakayama}},
  \bibinfo {author} {\bibfnamefont {T.}~\bibnamefont {Sato}}, \bibinfo {author}
  {\bibfnamefont {T.}~\bibnamefont {Takahashi}}, \bibinfo {author}
  {\bibfnamefont {M.}~\bibnamefont {Shi}},\ and\ \bibinfo {author}
  {\bibfnamefont {H.}~\bibnamefont {Ding}},\ }\href
  {https://doi.org/10.1103/PhysRevB.88.220508} {\bibfield  {journal} {\bibinfo
  {journal} {Phys. Rev. B}\ }\textbf {\bibinfo {volume} {88}},\ \bibinfo
  {pages} {220508} (\bibinfo {year} {2013})}\BibitemShut {NoStop}%
\bibitem [{\citenamefont {Richard}\ \emph {et~al.}(2015)\citenamefont
  {Richard}, \citenamefont {Qian},\ and\ \citenamefont
  {Ding}}]{RichardDing2015JPCM_APRES_IBS}%
  \BibitemOpen
  \bibfield  {author} {\bibinfo {author} {\bibfnamefont {P.}~\bibnamefont
  {Richard}}, \bibinfo {author} {\bibfnamefont {T.}~\bibnamefont {Qian}},\ and\
  \bibinfo {author} {\bibfnamefont {H.}~\bibnamefont {Ding}},\ }\href
  {https://doi.org/10.1088/0953-8984/27/29/293203} {\bibfield  {journal}
  {\bibinfo  {journal} {Journal of Physics: Condensed Matter}\ }\textbf
  {\bibinfo {volume} {27}},\ \bibinfo {pages} {293203} (\bibinfo {year}
  {2015})}\BibitemShut {NoStop}%
\bibitem [{\citenamefont {Avci}\ \emph {et~al.}(2012)\citenamefont {Avci},
  \citenamefont {Chmaissem}, \citenamefont {Chung}, \citenamefont {Rosenkranz},
  \citenamefont {Goremychkin}, \citenamefont {Castellan}, \citenamefont
  {Todorov}, \citenamefont {Schlueter}, \citenamefont {Claus}, \citenamefont
  {Daoud-Aladine}, \citenamefont {Khalyavin}, \citenamefont {Kanatzidis},\ and\
  \citenamefont {Osborn}}]{Avci2012PRB}%
  \BibitemOpen
  \bibfield  {author} {\bibinfo {author} {\bibfnamefont {S.}~\bibnamefont
  {Avci}}, \bibinfo {author} {\bibfnamefont {O.}~\bibnamefont {Chmaissem}},
  \bibinfo {author} {\bibfnamefont {D.~Y.}\ \bibnamefont {Chung}}, \bibinfo
  {author} {\bibfnamefont {S.}~\bibnamefont {Rosenkranz}}, \bibinfo {author}
  {\bibfnamefont {E.~A.}\ \bibnamefont {Goremychkin}}, \bibinfo {author}
  {\bibfnamefont {J.~P.}\ \bibnamefont {Castellan}}, \bibinfo {author}
  {\bibfnamefont {I.~S.}\ \bibnamefont {Todorov}}, \bibinfo {author}
  {\bibfnamefont {J.~A.}\ \bibnamefont {Schlueter}}, \bibinfo {author}
  {\bibfnamefont {H.}~\bibnamefont {Claus}}, \bibinfo {author} {\bibfnamefont
  {A.}~\bibnamefont {Daoud-Aladine}}, \bibinfo {author} {\bibfnamefont {D.~D.}\
  \bibnamefont {Khalyavin}}, \bibinfo {author} {\bibfnamefont {M.~G.}\
  \bibnamefont {Kanatzidis}},\ and\ \bibinfo {author} {\bibfnamefont
  {R.}~\bibnamefont {Osborn}},\ }\href
  {https://doi.org/10.1103/PhysRevB.85.184507} {\bibfield  {journal} {\bibinfo
  {journal} {Phys. Rev. B}\ }\textbf {\bibinfo {volume} {85}},\ \bibinfo
  {pages} {184507} (\bibinfo {year} {2012})}\BibitemShut {NoStop}%
\bibitem [{\citenamefont {Liu}\ \emph {et~al.}(2013)\citenamefont {Liu},
  \citenamefont {Tanatar}, \citenamefont {Kogan}, \citenamefont {Kim},
  \citenamefont {Lograsso},\ and\ \citenamefont {Prozorov}}]{Liu2013PRB}%
  \BibitemOpen
  \bibfield  {author} {\bibinfo {author} {\bibfnamefont {Y.}~\bibnamefont
  {Liu}}, \bibinfo {author} {\bibfnamefont {M.~A.}\ \bibnamefont {Tanatar}},
  \bibinfo {author} {\bibfnamefont {V.~G.}\ \bibnamefont {Kogan}}, \bibinfo
  {author} {\bibfnamefont {H.}~\bibnamefont {Kim}}, \bibinfo {author}
  {\bibfnamefont {T.~A.}\ \bibnamefont {Lograsso}},\ and\ \bibinfo {author}
  {\bibfnamefont {R.}~\bibnamefont {Prozorov}},\ }\href
  {https://doi.org/10.1103/PhysRevB.87.134513} {\bibfield  {journal} {\bibinfo
  {journal} {Phys. Rev. B}\ }\textbf {\bibinfo {volume} {87}},\ \bibinfo
  {pages} {134513} (\bibinfo {year} {2013})}\BibitemShut {NoStop}%
\bibitem [{\citenamefont {Liu}\ \emph {et~al.}(2014)\citenamefont {Liu},
  \citenamefont {Tanatar}, \citenamefont {Straszheim}, \citenamefont {Jensen},
  \citenamefont {Dennis}, \citenamefont {McCallum}, \citenamefont {Kogan},
  \citenamefont {Prozorov},\ and\ \citenamefont
  {Lograsso}}]{LiuProzorovLograsso2014PRB}%
  \BibitemOpen
  \bibfield  {author} {\bibinfo {author} {\bibfnamefont {Y.}~\bibnamefont
  {Liu}}, \bibinfo {author} {\bibfnamefont {M.~A.}\ \bibnamefont {Tanatar}},
  \bibinfo {author} {\bibfnamefont {W.~E.}\ \bibnamefont {Straszheim}},
  \bibinfo {author} {\bibfnamefont {B.}~\bibnamefont {Jensen}}, \bibinfo
  {author} {\bibfnamefont {K.~W.}\ \bibnamefont {Dennis}}, \bibinfo {author}
  {\bibfnamefont {R.~W.}\ \bibnamefont {McCallum}}, \bibinfo {author}
  {\bibfnamefont {V.~G.}\ \bibnamefont {Kogan}}, \bibinfo {author}
  {\bibfnamefont {R.}~\bibnamefont {Prozorov}},\ and\ \bibinfo {author}
  {\bibfnamefont {T.~A.}\ \bibnamefont {Lograsso}},\ }\href
  {https://doi.org/10.1103/PhysRevB.89.134504} {\bibfield  {journal} {\bibinfo
  {journal} {Phys. Rev. B}\ }\textbf {\bibinfo {volume} {89}},\ \bibinfo
  {pages} {134504} (\bibinfo {year} {2014})}\BibitemShut {NoStop}%
\bibitem [{\citenamefont {Tanatar}\ \emph {et~al.}(2010)\citenamefont
  {Tanatar}, \citenamefont {Ni}, \citenamefont {Bud’ko}, \citenamefont
  {Canfield},\ and\ \citenamefont {Prozorov}}]{Tanatar2010SST}%
  \BibitemOpen
  \bibfield  {author} {\bibinfo {author} {\bibfnamefont {M.~A.}\ \bibnamefont
  {Tanatar}}, \bibinfo {author} {\bibfnamefont {N.}~\bibnamefont {Ni}},
  \bibinfo {author} {\bibfnamefont {S.~L.}\ \bibnamefont {Bud’ko}}, \bibinfo
  {author} {\bibfnamefont {P.~C.}\ \bibnamefont {Canfield}},\ and\ \bibinfo
  {author} {\bibfnamefont {R.}~\bibnamefont {Prozorov}},\ }\href
  {https://doi.org/10.1088/0953-2048/23/5/054002} {\bibfield  {journal}
  {\bibinfo  {journal} {Superconductor Science and Technology}\ }\textbf
  {\bibinfo {volume} {23}},\ \bibinfo {pages} {054002} (\bibinfo {year}
  {2010})}\BibitemShut {NoStop}%
\bibitem [{\citenamefont {Bl\"ochl}(1994)}]{Blochl_PAW}%
  \BibitemOpen
  \bibfield  {author} {\bibinfo {author} {\bibfnamefont {P.~E.}\ \bibnamefont
  {Bl\"ochl}},\ }\href {https://doi.org/10.1103/PhysRevB.50.17953} {\bibfield
  {journal} {\bibinfo  {journal} {Phys. Rev. B}\ }\textbf {\bibinfo {volume}
  {50}},\ \bibinfo {pages} {17953} (\bibinfo {year} {1994})}\BibitemShut
  {NoStop}%
\bibitem [{\citenamefont {Kresse}\ and\ \citenamefont
  {Furthm\"uller}(1996)}]{VASP_ref2}%
  \BibitemOpen
  \bibfield  {author} {\bibinfo {author} {\bibfnamefont {G.}~\bibnamefont
  {Kresse}}\ and\ \bibinfo {author} {\bibfnamefont {J.}~\bibnamefont
  {Furthm\"uller}},\ }\href {https://doi.org/10.1103/PhysRevB.54.11169}
  {\bibfield  {journal} {\bibinfo  {journal} {Phys. Rev. B}\ }\textbf {\bibinfo
  {volume} {54}},\ \bibinfo {pages} {11169} (\bibinfo {year}
  {1996})}\BibitemShut {NoStop}%
\bibitem [{\citenamefont {Perdew}\ \emph {et~al.}(1996)\citenamefont {Perdew},
  \citenamefont {Burke},\ and\ \citenamefont
  {Ernzerhof}}]{perdew1996generalized}%
  \BibitemOpen
  \bibfield  {author} {\bibinfo {author} {\bibfnamefont {J.~P.}\ \bibnamefont
  {Perdew}}, \bibinfo {author} {\bibfnamefont {K.}~\bibnamefont {Burke}},\ and\
  \bibinfo {author} {\bibfnamefont {M.}~\bibnamefont {Ernzerhof}},\ }\href
  {https://doi.org/10.1103/PhysRevLett.77.3865} {\bibfield  {journal} {\bibinfo
   {journal} {Phys. Rev. Lett.}\ }\textbf {\bibinfo {volume} {77}},\ \bibinfo
  {pages} {3865} (\bibinfo {year} {1996})}\BibitemShut {NoStop}%
\bibitem [{\citenamefont {Kresse}\ and\ \citenamefont
  {Hafner}(1993)}]{VASP_ref}%
  \BibitemOpen
  \bibfield  {author} {\bibinfo {author} {\bibfnamefont {G.}~\bibnamefont
  {Kresse}}\ and\ \bibinfo {author} {\bibfnamefont {J.}~\bibnamefont
  {Hafner}},\ }\href {https://doi.org/10.1103/PhysRevB.47.558} {\bibfield
  {journal} {\bibinfo  {journal} {Phys. Rev. B}\ }\textbf {\bibinfo {volume}
  {47}},\ \bibinfo {pages} {558} (\bibinfo {year} {1993})}\BibitemShut
  {NoStop}%
\bibitem [{\citenamefont {Oen}(1973)}]{Oen1973}%
  \BibitemOpen
  \bibfield  {author} {\bibinfo {author} {\bibfnamefont {O.~S.}\ \bibnamefont
  {Oen}},\ }\href {https://doi.org/10.2172/4457758} {\emph {\bibinfo {title}
  {Cross sections for atomic displacements in solids by fast electrons}}},\
  \bibinfo {type} {Tech. Rep.}\ (\bibinfo  {institution} {Office of Scientific
  and Technical Information ({OSTI}), {ID}: 4457758, Report Num.: ORNL-4897},\
  \bibinfo {year} {1973})\BibitemShut {NoStop}%
\end{thebibliography}%

\end{document}